# Artificial Intelligence and Diabetes Mellitus: An Inside Look Through the Retina

Yasin Sadeghi-Bazargani [1, #], Majid Mirzaei [1, #], Navid Sobhi [2, #], Mirsaeed Abdollahi [2], Ali Jafarizadeh [2, *], Siamak Pedrammehr [3], Roohallah Alizadehsani [4, *], Ru-San Tan [5,6], Sheikh Mohammed Shariful Islam [7,8,9], U. Rajendra Acharya [10,11]

[1] *Student Research Committee, Tabriz University of Medical Sciences, Tabriz, Iran*

[2] *Nikookari Eye Center, Tabriz University of Medical Sciences, Tabriz, Iran*

[3] *Faculty of Design, Tabriz Islamic Art University, Tabriz, Iran*

[4] *Institute for Intelligent Systems Research and Innovation (IISRI), Deakin University, VIC 3216, Australia*

[5] *National Heart Centre Singapore, Singapore*

[6] *Duke-NUS Medical School, Singapore*

7 *Institute for Physical Activity and Nutrition, School of Exercise and Nutrition Sciences, Deakin University, Geelong, VIC, Australia*

8 *Cardiovascular Division, The George Institute for Global Health, Newtown, Australia*

9 *Sydney Medical School, University of Sydney, Camperdown, Australia*

10 *School of Mathematics, Physics, and Computing, University of Southern Queensland, Springfield, QLD 4300, Australia*

11 *Centre for Health Research, University of Southern Queensland, Australia*

# *Contributed equally*

\***Corresponding Authors**:

Ali Jafarizadeh, MD, MPH

    *Nikookari Eye Center, Tabriz University of Medical Sciences, Tabriz, Iran*

    *Tel: +98 901 098 0062*

    *Postal code:* 51666/14766

    *Email: Jafarizadeha@tbzmed.ac.ir, Ali.jafarizadeh.md@gmail.com*

    https://orcid.org/0000-0003-4922-1923

Roohallah Alizadehsani, PhD

    *Institute for Intelligent Systems Research and Innovation (IISRI), Deakin University, VIC 3216, Australia*

    *Tel: +61 3 524 79394*

    *Postal: 75 Pigdons Rd, Waurn Ponds VIC 3216, Australia*

    *Email: r.alizadehsani@deakin.edu.au*

    https://orcid.org/0000-0003-0898-5054

**Abstract**

Diabetes mellitus (DM) predisposes patients to vascular complications. Retinal images and vasculature reflect the body's micro- and macrovascular health. They can be used to diagnose DM complications, including diabetic retinopathy (DR), neuropathy, nephropathy, and atherosclerotic cardiovascular disease, as well as forecast the risk of cardiovascular events. Artificial intelligence (AI)-enabled systems developed for high-throughput detection of DR using digitized retinal images have become clinically adopted. Beyond DR screening, AI integration also holds immense potential to address challenges associated with the holistic care of the patient with DM. In this work, we aim to comprehensively review the literature for studies on AI applications based on retinal images related to DM diagnosis, prognostication, and management. We will describe the findings of holistic AI-assisted diabetes care, including but not limited to DR screening, and discuss barriers to implementing such systems, including issues concerning ethics, data privacy, equitable access, and explainability. With the ability to evaluate the patient's health status vis a vis DM complication as well as risk prognostication of future cardiovascular complications, AI-assisted retinal image analysis has the potential to become a central tool for modern personalized medicine in patients with DM.

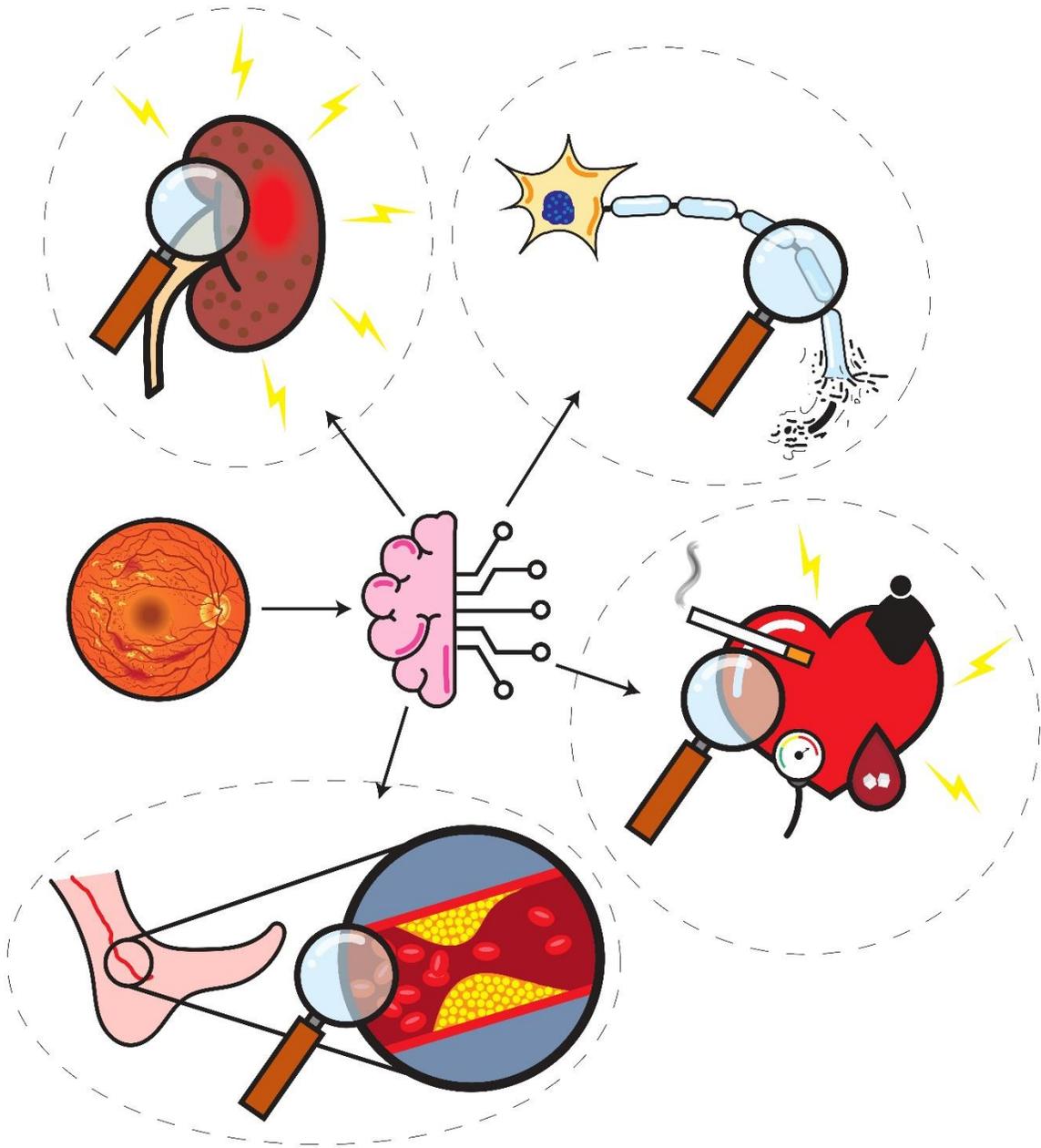

**Figure 1**. An illustration depicts the various diabetic complications diagnosed by retinal images using AI. (Graphical abstract)

1. **Introduction**

Diabetes mellitus (DM) affects more than 422 million people globally and is a significant cause of mortality and morbidity [1,2]. It predisposes patients to macrovascular complications — i.e., atherosclerotic cardiovascular diseases like coronary heart disease, cerebrovascular disease, and peripheral arterial disease (the risk of death from cardiovascular diseases is two to six times greater in diabetic vs. non-diabetic populations [1]) — as well as microvascular complications like peripheral neuropathy [3], nephropathy (end-stage kidney disease is ten times more prevalent in diabetic vs. non-diabetic populations [4]), and retinopathy.

With a prevalence of 34.6% among diabetic patients [5], diabetic retinopathy (DR) is the leading cause of preventable adult blindness [6]. DR can be categorized into non-proliferative and proliferative stages; the incidence of sight-threatening diabetic macular edema (DME), which can occur in either stage, rises as DR progresses [6]. With the rising prevalence and incidence of DM, the burden of DR and consequent vision loss continues to increase. Globally, 1.4 million patients with DM have severe non-proliferative DR and proliferative DR and will require treatment to arrest or retard vision deterioration [5]. Preemptive systematic and continual screening of asymptomatic diabetic patients, with early initiation of anti-proliferative treatment where indicated, is obligatory for preventing the clinical sequelae and reducing the human costs of DR [7].

Fundoscopic retinal examination is traditionally used in the clinic to screen for DR. The retinal vasculature is the only part of the human body's microcirculation that can be visualized non-invasively [8]. The retina is a veritable "window" to the state of health of the microcirculation as a whole—shedding light not only on DR but also on other diabetic microvascular complications like neuropathy and nephropathy [9]—as well as the risk of macrovascular disease. The introduction of digital fundal photography, which enables offline analysis of retinal images by experts, has contributed to the successful implementation of national DR screening programs at scale. With the advent of artificial intelligence (AI) applications in medicine, especially in the interpretation of digital medical images (which are by nature objective and accessible), many AI-enabled systems developed for high-throughput detection of DR using digitized retinal images have become clinically adopted [10, 11].

Beyond DR screening, AI integration also holds immense potential to address challenges associated with the holistic care of the patient with DM. As alluded to above, the retinal vasculature (and image) contains diagnostic and prognostic clues to the micro- and macrovascular health of the whole body. In contrast to existing review studies centered solely on diabetic retinopathy within the realm of diabetes, retina, and artificial intelligence, our study takes a broader approach [12-20]. In this work, we aim to comprehensively review the literature for AI applications based on retinal images related to DM diagnosis, prognostication, and management. We believe the study will offer insights into applications of AI technology for diabetes care, including but not limited to DR screening. In addition, we will discuss barriers to implementing such systems, including data privacy concerns, algorithm transparency, regulatory compliance, ethical considerations surrounding decision-making autonomy, and equitable access to technology-enabled healthcare solutions.

## 2. Literature search method

We performed a comprehensive search of PubMed, Medline, Google Scholar, Scopus, Web of Sciences, and IEEE Xplore databases for scholarly articles published in the English language up to October 1, 2023, using various combinations (by using Boolean operators: AND, OR, NOT) of the following main keywords: "diabetes mellitus," "retina," "artificial intelligence."

Other keywords are "diabetic care," "diabetes complications," "diabetic foot," "diabetic nephropathy," "diabetic neuropathy," "diabetic retinopathy," "diabetic macular edema," "cardiovascular risk assessment," "peripheral arterial disease," "fundus image," "personalized medicine," "diagnosis," "prognosis," "deep learning," "machine learning," "prediction," and "machine vision" [21].

In selecting articles for inclusion in our study, we employed a comprehensive approach focused on identifying original studies with clear research goals, meticulous methodology, and extensive technical details. In our evaluation process, we defined 'technical details' to include the technological aspects and analytical methods used in the studies. This entailed examining the types of artificial intelligence algorithms deployed, data processing techniques, specific retinal imaging methods, and any advanced computational strategies implemented. We also discussed the software applications, statistical techniques, and validation procedures in these research studies. Our focus on these technical details aimed to ensure a comprehensive assessment of the methodologies and technological advancements in AI applications in diabetic retinal image analysis.

## 3. Results and discussion

In this investigation, we conducted a comprehensive and discerning examination of the academic literature, particularly emphasizing seminal original articles published in leading Q1 journals (based on scimagojr.com). The aim was to gain a comprehensive understanding of the incorporation of artificial intelligence in diabetes management, focusing on retinal image analysis. Our review was strategically centered on 29 pivotal original studies that have significantly impacted the advancements in diabetic care through the application of AI.

The search strategy yielded 2716 research results, from which we meticulously eliminated 1089 duplicates to refine the scope of our examination. Subsequently, a preliminary evaluation of titles and abstracts was conducted, excluding 1376 studies that did not closely align with our focused criteria on AI's utility in diabetes management through retinal imaging. The remaining articles underwent a rigorous assessment of the complete text, excluding 179 papers due to either deviation from our research objectives or methodological limitations. This detailed process resulted in the selection of 72 scholarly articles chosen for their clear research goals, meticulous methodology, and extensive technical details. The article selection process, as illustrated in Figure 2, was precise, ensuring the inclusion of only the most relevant and rigorous studies. Case reports and animal studies were excluded to maintain the integrity and focus of our examination. Our research team carefully checked every paper chosen for this study. Our team rigorously evaluated each selected paper, scrutinizing the title and abstract to verify their relevance to the key aims of our research.

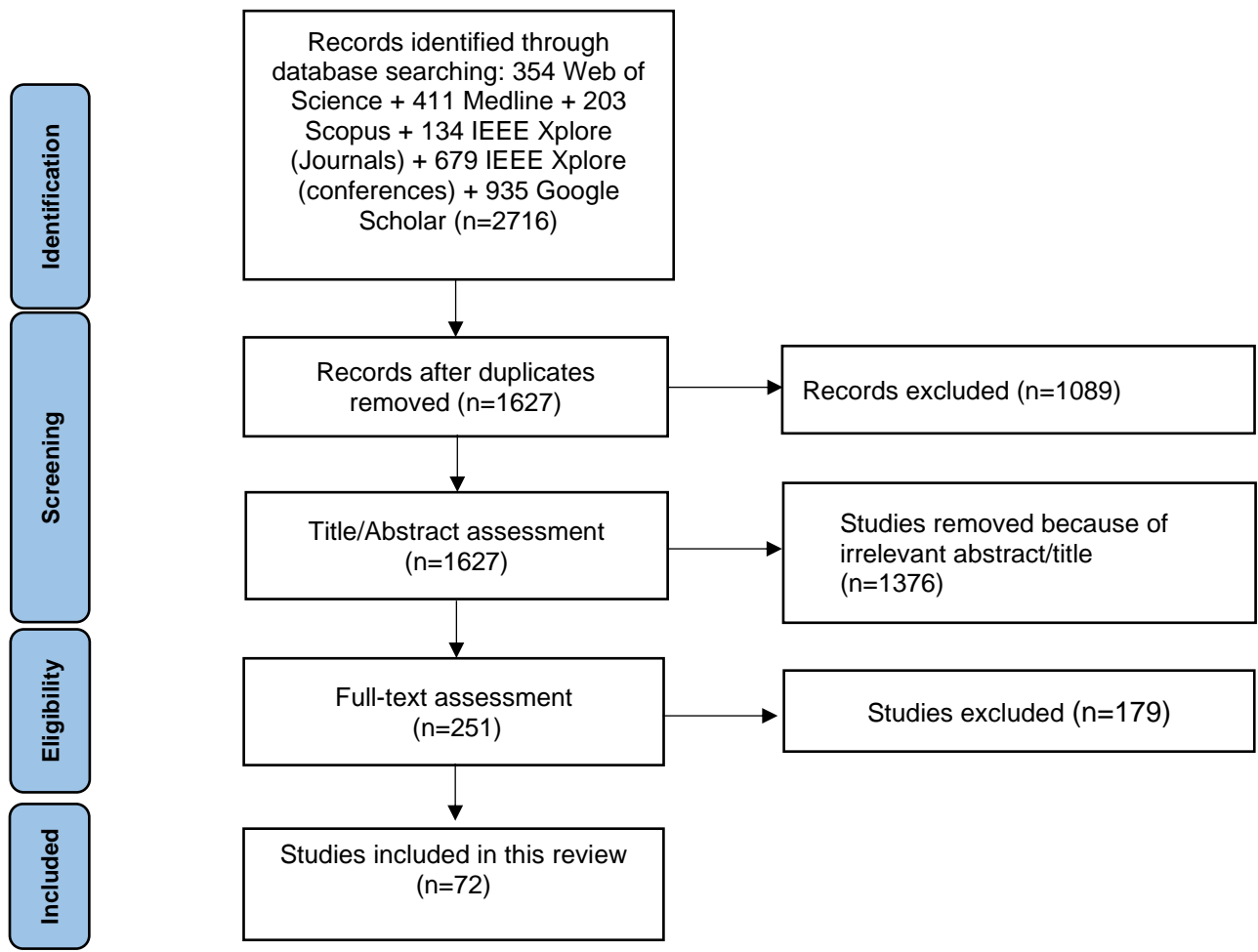

**Figure 2.** Preferred Reporting Items for Systematic Reviews and Meta-Analyses (PRISMA) flow diagram.

## 3.1 Diabetic retinopathy and diabetic macular edema

### 3.1.1 DR screening

In the landmark study by Abramoff et al., an AI system based on supervised machine learning (ML) with logistic regression attained 96.8% sensitivity and 59.4% specificity for detection of referable DR. Despite the modest specificity, the system was projected to halve the burden compared to manual screening by experts [22]. Using convolutional neural network (CNN) architectures inspired by AlexNet and VGGNet, the same team developed a new deep learning (DL) model that demonstrated improved 87% specificity for detecting referable DR without loss of sensitivity [23]. EyeArt and Retmarker, two AI systems that reported 93.8% and 85% sensitivities, respectively, for referable DR, were found to be associated with the lowered cost of DR screening [24]. Other economic modeling studies also showed that semi-automated and fully automated screening methods by human experts are more cost-effective than traditional manual screening [25, 26]. IDx-DR, the first medical device implementing AI for detecting more than mild DR, was approved by the FDA in 2018 [27].

In low-income countries with limited healthcare resources, the lack of established DM screening and programs for blood glucose control results in higher incidences of diabetes complications. In one study conducted in sub-Saharan Africa, the rate of progression from no DR to sight-threatening DR was five times higher than in Europe [28, 29]. Despite this, universal annual DR screening for all patients with DM remains an elusive goal [29]. The solution may come in the form of technological advances. AI analysis of digital fundal photographs can potentially ameliorate the cost of nationwide screening programs with comparable or superior diagnostic performance [5, 30, 31]. Whether models developed and validated in other countries with different ethnic mixes, apply to local populations is a logical concern. In study by Bellemo et al. [32], an ensemble model comprising adapted VGGNet and ResNet architectures trained on a Singapore database of color retinal images was demonstrated to have acceptable diagnostic performance—0.973 receiver operating characteristic area-under-curve (AUC) for referable DR; 99·42% sensitivity for sight-threatening DR; 97·19% sensitivity for diabetic macular edema (DME)—in a real-world DM population in Zambia. Separately, a real-world prospective interventional cohort study in Thailand reported the accuracy of a DL-based system in detecting sight-threatening DR to be 94.7%, compared to 93.5% for human experts [33]. Among novel cost-saving approaches, retinal images can be acquired using smartphone-based retinal cameras. This method was demonstrated to be feasible in a study using EyeArt (ensemble of deep artificial neural networks) AI software, which attained 99.1% sensitivity and 80.4% specificity for detecting sight-threatening DR based on smartphone retinal images (with online processing) analyzed by AI vs. human experts [34].

### 3.1.2 Advanced DR screening for DME

DME, an accumulation of fluid in the macula, can lead to visual loss at any stage of DR. DME at any stage of DR mandates referral to a specialist. The definitive diagnosis requires either macular thickness measurement using optical coherence tomography (OCT) or visualization of edema on fluorescein angiography. Hard exudates within one optic disc diameter on the color fundus photograph (CFP) are a surrogate for DME [35], with a significant false positive rate of up to 42% in the UK screening program [36]. The Danish guidelines recommend OCT as a second-line screening method for DR [37].

AI plays a significant role in diagnosing and classifying DME [38]. A DL model trained on fundus photographs attained an AUC of 0.89 for discriminating center-involved DME, with superior specificity and positive predictive value vs. human experts [39]. More DL systems have been developed for the automatic detection of DME on OCT images that can also discern the grades of severity, ranging from diffuse retinal thickening cystoid macular edema to serous retinal detachment [40]. A meta-analysis of 53 studies reported 96% and 94% sensitivities for AI-assisted DME detection based on OCT and fundus images, respectively [41]. Another meta-analysis of 19 published DL algorithms reported pooled 96.0% (95% CI: 93.9% to 97.3%) sensitivity and 99.3% (95% CI: 98.2% to 99.7%) specificity for DME detection using OCT images [42]. In a comparison of two DL models, the Optic-Net model (98% accuracy, 100% specificity) outperformed Dense-Net (94% accuracy, 96% specificity) for OCT-based DME classification [43]. Rui Liu et al. [44] trained a faster R-CNN model with ResNet101 backbone on more than 50,000 labeled fundus images and 20,000 OCT B-scans acquired from patients from multiple centers and reported excellent 97.78% sensitivity, 98.38% specificity, and 0.981 AUC for detection of referable DR; and 91.30% sensitivity, 97.46% specificity, and 0.944 AUC for detection of DME. The incremental diagnostic utility of OCT-based AI analysis—combined CFP plus OCT screening-detected cases of DME that would have been missed on CFP analysis alone—lends support for adding OCT to standard CFP-based AI screening programs. However, the cost-effectiveness should be further assessed.

AI evaluation of OCT angiography, which generates angiographic images of retinal vessels from motion imaging of volumetric blood flow signals without contrast, has recently attracted attention. Ryu et al. [45] compared two models to detect early DR using OCT angiography: the first was an ML-based classifier comprising segmentation (using U-Net, which was able to distinguish blood vessels and the foveal avascular zone in the OCT angiography image), feature extraction, and classification; and the second, a CNN-based classifier that processed OCT angiography images directly using a ResNet101 architecture. The latter attained excellent 91-98% accuracy, 86-97% sensitivity, 94-99% specificity, and 0.919-0.976 AUC, which compared favorably against ultra-widefield fluorescein angiography.

### 3.1.3 Grading of DR severity

Classification of CFP images into various grades of DR severity (Figure 3) based on retinal vascular changes—no retinopathy, mild non-proliferative DR, moderate non-proliferative DR, severe non-proliferative DR, and proliferative DR [20]—may offer additional insights into disease progression and prognosis [46]. This is conventionally accomplished via manual fundal examination by experts DR [47], a scarce resource and not readily accessible. Promising AI-based tools can categorize DR grades, reducing healthcare costs and burdens [48, 49]. Gulshan et al. [50] developed a

CNN model with Inception-v3 architecture and transfer learning that outputs five independent binary classifiers for DR grading. The model attained good performance for grading DR: 84.0% (95% CI, 75.3%-90.6%) sensitivity and 98.8% (95% CI, 98.5%-99.0%) specificity for detection of severe or worse DR; and 90.8% (95% CI, 86.1%-94.3%) sensitivity and 98.7% (95% CI, 98.4%-99.0%) specificity for detection of DME. Takahashi et al. [51] used GoogLeNet CNN to grade DR using four different 45-degree field CFP images captured per eye. The network attained 81% mean accuracy and prevalence-adjusted bias-adjusted kappa of 0.74 (i.e., correct answer with maximum probability in 402 of 496 CFP images), outperforming standard manual grading by experts based on one CFP image. Other researchers also reported high sensitivity and specificity rates for DR grading using various models [52-54].

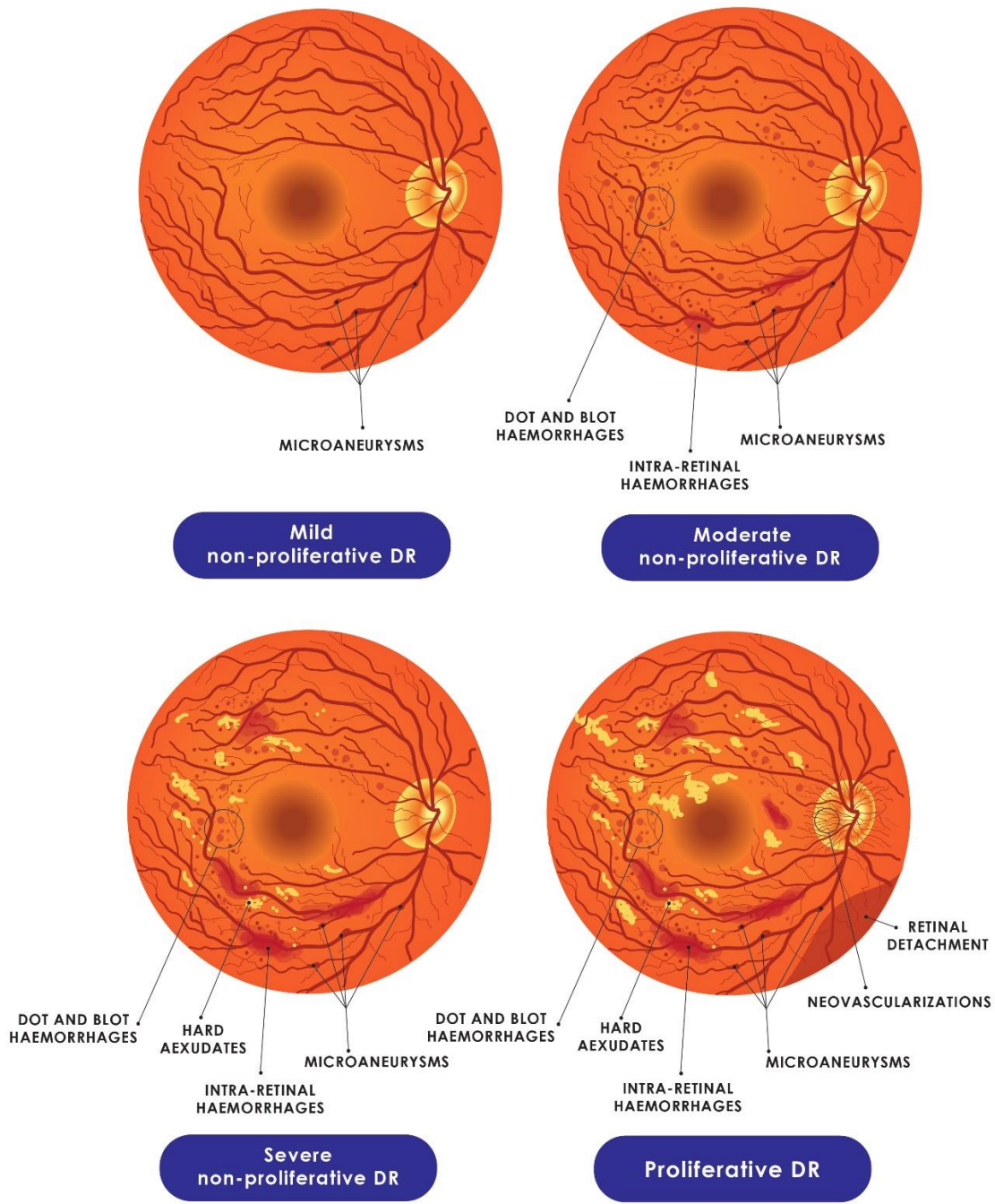

**Figure 3.** Schematic representation of fundoscopy showing the staging of diabetic retinopathy.

On fundoscopy, DR-induced vascular changes exhibit heterogeneous distribution; manual analysis typically assigns the highest observed class to the entire image. With AI analysis of the CFP image, the basis for severity grading is not apparent, which limits the explainability of AI decision-making. To address this, some researchers have developed DL neural networks for lesion-based classification [50, 54, 55]. Wang et al. [56] trained a CNN-based DR lesion classifier, Lesion-Net, on 12,252 fundus images of patients with DM. The model contains two branches: the inferior uses a fully convolutional network (FCN-32s) for lesion segmentation predictions of eight types of common DR lesions to diagnose and stage DR, and the superior branch integrates convolutional layers from Inception-v3 architecture to classify referable vs. non-referable DR classes. Lesion-Net attained a good 0.943 AUC, 90.6% sensitivity, and acceptable 80.7% specificity for five-stage DR grading, as well as an improved AUC of 0.936 vs. 0.928 for the internal test set and 0.977 vs. 0.964 for the external test set for binary referable vs. non-referable DR classification. According to a recent study, the Inception-v3 model, trained on a dataset of 35,126 retinal images, delivered the best performance for DR diagnosis among ML/DL models, yielding an accuracy rate of approximately 97% in feature extraction [57]. Concurrently, the MobileNetV3 model has been identified as optimal for data classification tasks in the same domain, achieving an accuracy of 98.56% [58].

Sandhu et al. combined OCT and OCT angiographic images with primary clinical and demographic data collected from 111 patients to train an AI model for screening and staging DR [59]. They developed a novel computer-aided design system to grade non-proliferative DR into mild vs. moderate stages. They reported 98.7% accuracy, 100% sensitivity, 97.8% specificity, 99% differential scanning calorimetry (DSC), and 0.981 AUC (progressive improvements in almost all metrics were observed as OCT angiography, clinical, and demographic data were incrementally added to the model). Wang et al. [60] used ultra-widefield fluorescein angiographic images of 399 patients to train an AI model to differentiate normal, non-proliferative DR, and proliferative DR. The developed model attained 88.50% accuracy.

**3.1.4 Individualized DR screening**

With annual DR screening, about 11% of DM patients with mild non-proliferative DR in both eyes progress to sight-threatening DR every year [61]. Given increasing DM incidence and anticipated DM-related resource utilization, individualized screening with varying intervals may be more cost-effective as an alternative to routine annual DR screening. Aspelund et al. used six parameters—sex, DM type, DM duration, glycated hemoglobin, blood pressure, and DR grade of DR—to statistically model DR screening intervals [62-64]. Compared with routine annual DR screening, they and other groups reported 40-60% reductions in screening visits and 11-40% increases in patients developing sight-threatening DR at the next screening visit [62-64]. AI can potentially improve upon the accuracy of predictions of such mathematical models. Piri et al. analyzed demographic, laboratory, and comorbidity data of more than 1.4 million DM patients using supervised ML models and artificial neural networks and were able to predict the presence of DR with an accuracy of 92.76% [64]. These observations suggest that risk stratification of asymptomatic DM without retinal imaging is feasible and can potentially be used to rationalize the timing of initiation and the intensity of the frequency of DR screening, especially in resource-limited healthcare systems.

**3.1.5 Individualized DR follow-up**

There is considerable inter-individual variability in the progression of DR, even among DM patients diagnosed with the same DR grade. Post-DR diagnosis, the frequency of follow-up visits, the timing of treatment initiation, and the treatment time frame should ideally be tailored to the individual [65]. Skevofilakas et al. [66] developed a decision support system for assessing DR progression using clinical data collected from the electronic health records of 55 type I DM patients who were followed over five years. The system used strongly linked risk factors—age, DM duration, glycated hemoglobin, cholesterol, triglyceride, hypertension incidence rate, and DM treatment duration—to predict DR progression, attaining high 97% accuracy with receiver operating characteristic analysis on the most accurate primary classifier, and higher overall decision support system sensitivity, specificity, and accuracy. Arcadu et al. [67] developed an algorithm for predicting worsening DR using 7-field CFPs acquired from DM patients monitored over two years. Baseline CFPs were used to train the model to indicate the presence of 2-step or more worsening of DR on the Early Treatment Diabetic Retinopathy Scale during the follow-up visits. Based on an Inception-v3 architecture, separate deep CNNs were trained in parallel for each of the seven fields via transfer learning initialized with ImageNet weights, and the generated probabilities were consolidated using random forests. The trained DL model predicted DR worsening at 6, 12, and 24 months with AUCs of $0.68 \pm 0.13$ (sensitivity, $66\% \pm 23\%$; specificity $77\% \pm 12\%$), $0.79 \pm 0.05$ (sensitivity, $91\% \pm 8\%$; specificity, $65\% \pm 12\%$), and $0.77 \pm 0.04$ (sensitivity, $79\% \pm 12\%$; specificity, $72\% \pm 14\%$), respectively, which lends support to the use of AI to refine and personalize prognostication even among individuals within the same DR grade as assessed by traditional manual methods. Of note, the aggregate result was superior to the individual field-specific deep CNNs, which indirectly validates the choice of Inception-v3 design. In particular, the peripheral parts of the retina contributed more to the predictions than central regions, with model performance dropping significantly when peripheral zones were excluded. Further, the attribution map showed that the model could differentiate among microaneurysms, hemorrhages, and hard exudates, laying the groundwork for potential lesion-specific analysis and predictions.

**3.1.6 DR treatment: Who to treat and how to treat?**

Intravitreal injection of anti-vascular endothelial growth factor (VEGF) medications, like ranibizumab, bevacizumab, and aflibercept, is an indicated treatment for sight-threatening DR, especially DME [68-71]. OCT is often used to monitor the therapeutic response. Through analysis of OCT images, AI models can predict individual patient responses to anti-VEGF therapy and potentially facilitate personalized treatment approaches for DME [72]. Advanced AI algorithms have been developed that can evaluate OCT parameters—central macular fluid volume, integrity of the ellipsoid zone, intraretinal fluid, subretinal fluid, hyperreflective retina foci, and external limiting membrane [73]—to predict visual acuity trajectories in DME, thereby providing clinicians with objective parameters for DME diagnosis and follow-up. Liu et al. [74] combined DL and classical machine learning (CML) models. They trained AlexNet, VGG16, ResNet18, and an ensemble of the three DL architectures on a dataset of 304 pre-treatment OCT images of patients with DME. Fifteen OCT features generated by the DL ensemble model were then used to train conventional ML algorithms—Lasso, support vector machine, decision tree, and random forest—to predict post-treatment central foveal thickness and best-corrected visual acuity values at one month after three months of anti-VEGF injections [74]. However, the model failed to accurately predict the post-treatment values, suggesting that OCT images alone were insufficient as sole model inputs. Gallardo et al. [75] developed an ML system to assess the burden of anti-VEGF treatment—defined as low, moderate, and high based on the interval of injections—in a treat-and-extend regimen for DME and retinal vein occlusion using demographic data and OCT images obtained patients at two consecutive clinic visits [75]. The proposed random forest-based supervised ML model to predict the

1-year treatment requirement with a reasonable AUC so that the decision-making process was rendered interpretable. In particular, all features related to intra-retinal fluid were important for predicting low vs. high treatment demand.

Besides predicting treatment response, AI can be applied to direct planning of the therapeutic procedure. Focal or grid laser photocoagulation—during which a series of controlled photocoagulations are delivered to retinal areas with pathology [76, 77]—is another indicated treatment for DME and proliferative DR. By normalizing oxygen partial pressures in peripheral avascular regions of the retina, the treatment induces regression of newly formed vessels. It facilitates the rates of formation of aberrant vessels, vitreous hemorrhage, and membranes. Treatment efficacy highly depends on the siting and dosing of the administered photocoagulates [78]. Standard pre-determined patterns for photocoagulation cannot account for individual differences in the shapes and patterns of macular edema and anatomical variations of retinal vasculature [76, 79]. Furthermore, manually mapping a coagulation pattern requires surgical expertise and considerable time costs [79]. AI can be harnessed to automate retinal segmentation such that only pre-determined areas of the retina are coagulated, thereby increasing the precision of laser photocoagulation and minimizing unwanted side effects. Personalized, high-quality coagulation maps were generated by inputting patient information into a novel AI software. The resultant enhanced precision in the localization of the exact point of burn and control of the amount of power delivered compared with manual methods resulted in a nine-fold decrease in the probability of laser burns beyond the borders of the edema, as well as shortened procedural preparation time and reduced postoperative complications [79].

**3.2 Beyond the eye: Systemic micro- and macrovascular complications**

Table 1 summarizes studies of AI models developed using retinal images to detect DM complications or sequelae outside the eye. The explanations are detailed in the following sections

**3.2.1 Predicting Cardiovascular Risk Using Retinal Photographs**

DM increases the risk of cardiovascular events by 2 to 4 folds [1, 80]; cardiovascular diseases constitute the most common cause of death in DM. Cardiovascular risk can be calculated from clinical (e.g., age, sex, body mass index, blood pressure, etc.) and laboratory data (e.g., glycated hemoglobin, lipids, creatinine clearance, etc.). Based on the same information, AI models can modestly improve performance, raising the AUC from 0.69 to 0.75 [81]. Of note, such models are not specific to the DM population. In DM patients, retinal parameters have been shown to correlate with cardiovascular outcomes independent of the underlying risk factors [82]. The abundance of retinal image databases from patients with DR has enabled researchers to extract new associations from these databases using AI. Poplin et al. trained an Inception-v3 model on 284,335 retinal images and reported 0.70 AUC for image-based prediction of major cardiovascular events [83]. The accuracy of cardiovascular risk prediction can be enhanced using combined retinal images and accessible clinical and demographic data [84]. While preliminary, these findings are promising and warrant research into cardiovascular risk assessment as part of DR screening in diabetic patients.

**3.2.2 Beyond the optic nerve: Predicting diabetic neuropathy**

Like DR, diabetic neuropathy is a common microvascular complication of DM. It causes skin ulcers, increases the risk of limb amputation, and reduces the quality of life, accounting for 27% of DM-related healthcare costs [85-87]. Routine annual screening via clinical neurological examination is insensitive and often fails to detect early changes before irreversible damage [88]. Retinal vasculature, which reflects the systemic microcirculation, may hold clues to the development of diabetic neuropathy as well. Benson et al. used a pre-trained VGG16 CNN and a support vector machine classifier to analyze retinal images. They attained 89% accuracy, 78% sensitivity, and 95% specificity for detecting physician-diagnosed diabetic peripheral neuropathy (which is subjective, insensitive, and time-consuming to diagnose clinically) [89]. Among DL methods for diagnosing diabetic neuropathy on CFP, the most effective are SqueezeNet, Inception, and DenseNet, which yielded the best AUCs of 0.8013 (±0.0257) and 0.7097 (±0.0031) during validation and testing, respectively, when evaluated on various datasets, with and without pre-trained weights [90].

### 3.2.3 Eyes on the kidneys: Screening for chronic kidney disease

Diabetic nephropathy, another microvascular complication of DM, is a common cause of chronic kidney disease (CKD) [91-94]. The global prevalence of CKD is 10-14% [95-97], with nearly a million deaths attributed to CKD in 2013, of which a fifth were caused by DM nephropathy [98, 99]. Early diagnosis by blood and urine screening assays is recommended to select at-risk patients for initiation of treatment to retard kidney function deterioration. Retinal images hold clues to systemic microcirculatory health and may be used to screen for diabetic nephropathy. In a study by Zhang et al. [100], clinical parameters and retinal images were used separately and combined to predict diabetic nephropathy in a CondenseNet DL model. The model attained AUCs of 0.916, 0.911, and 0.938 for clinical risk factors, retinal images, and hybrid inputs, respectively. In a similar study, a ResNet50 DL model attained AUCs of 0.861, 0.918, and 0.930 for clinical risk factors, retinal images, and hybrid inputs, respectively [101]. These findings corroborate the superior performance of AI-based diabetic nephropathy detection using combined clinical and retinal image inputs; retinal screening alone may not suffice [102].

### 3.2.4 Retinal images for diagnosing peripheral arterial disease

Peripheral arterial disease (PAD) is a macrovascular complication of DM, with an incidence rate of 25-30% [103]. The elevated risk of PAD in type 2 DM is attributable to the interplay of traditional cardiovascular risk factors—age, sex, race, smoking, pulse pressure, glycated hemoglobin, albuminuria, and hyperlipidemia [104]—as well as diabetes-specific elements, including postprandial hyperglycemia, advanced glycation end-products, lipoproteins, and hypercoagulability [105]. Low ankle-brachial pressure index is commonly used to screen for lower limb PAD non-invasively but may be falsely high in elderly people with inelastic arteries. Alternative noninvasive diagnostic techniques are ultrasound Doppler waveform analysis and toe-brachial index [106]. A deep neural network architecture has recently been demonstrated to be promising for detecting PAD using CFP. The most successful model in this inventive methodology attained a ROC AUC score of 0.890. Notably, visualizing the attention weights used by the network provides valuable insights into its decision-making process, particularly highlighting the significance of ocular features in PAD. Statistical analysis of the model's performance confirmed that the optic disc and the temporal arcades are assigned significantly higher importance ($p < 0.001$) than the retinal background in the detection process. These results robustly support the feasibility and effectiveness of utilizing modern DL methodologies to detect PAD [107]. However, a major challenge to AI diagnosing PAD on retinal images is the difficulty of integrating high-quality images and comprehensive training datasets with labeled PAD cases into the AI frameworks [107, 108]. Figure 4 depicts a systematic overview of the use of AI in diabetes care.

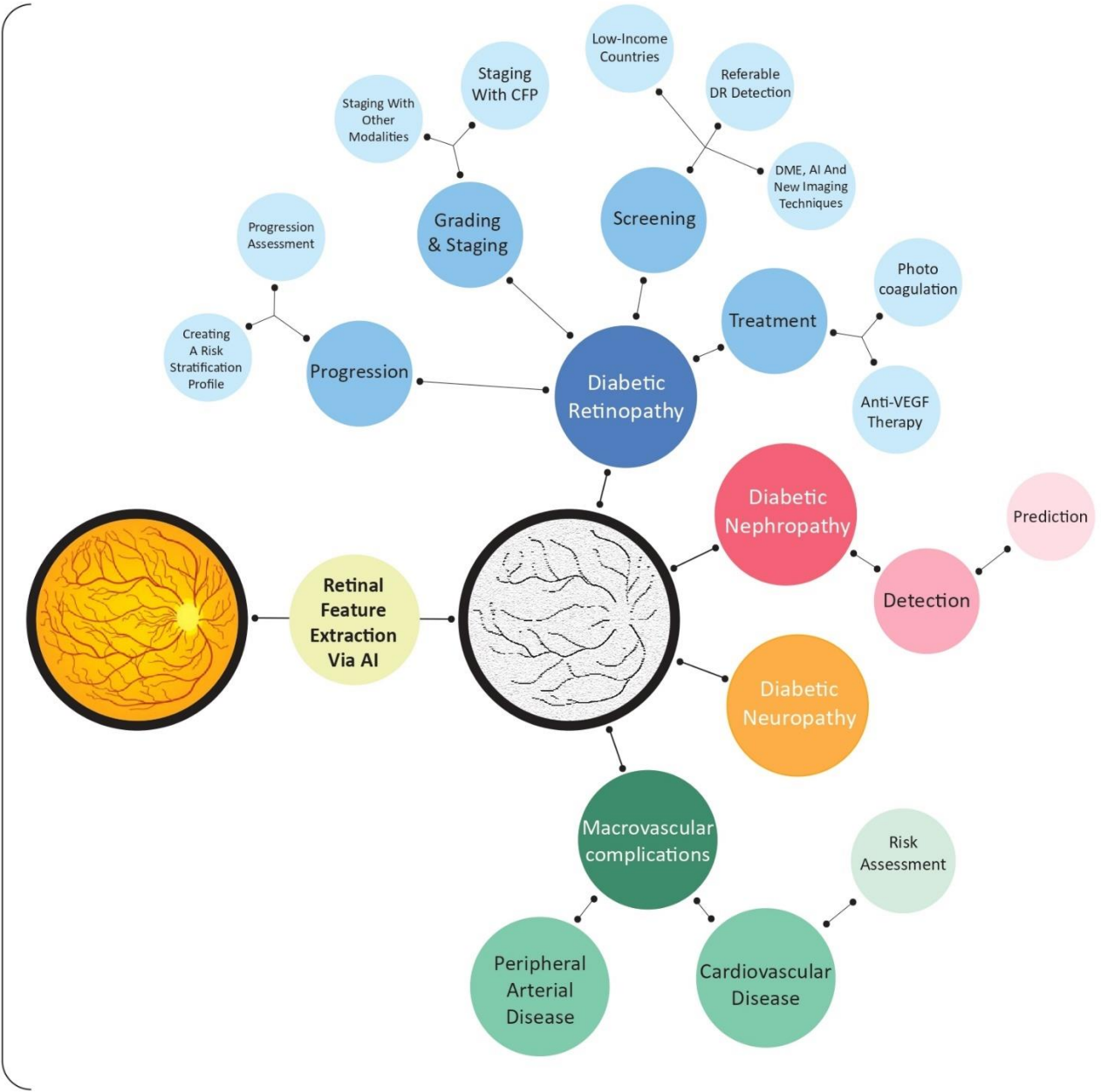

**Figure 4.** The diagram provides a structured overview of the integration of AI in diabetic care, showing the pathways where AI has significant applications, including diabetic retinopathy, nephropathy, neuropathy, cerebrovascular disease, peripheral arterial disease, and cardiovascular assessments. Each branch highlights the potential of AI in enhancing diagnosis, monitoring, and management within these specialized areas of diabetic care.

Table 1. Summary of Performance Metrics of Models and Architectures for diabetes complications diagnosis

| Study | AI modeling algorithms | Architecture | clinical predictors | Metric | Results |
|---|---|---|---|---|---|
| **Diabetic Cardiovascular disease** | | | | | |
| **Poplin [83]** | DL | Inception-v3 | Age | R2 (95% CI) | 0.74 (0.73,0.75) |
| | | | Gender | AUC (95% CI) | 0.97 (0.966,0.971) |
| | | | Smoking status | AUC (95% CI) | 0.71 (0.70,0.73) |
| | | | BMI | R2 (95% CI) | 0.13 (0.11,0.14) |
| | | | SBP | R2 (95% CI) | 0.36 (0.35,0.37) |
| | | | DBP | R2 (95% CI) | 0.32 (0.30,0.33) |
| | | | HbA1c | R2 (95% CI) | 0.09 (0.03,0.16) |
| **Diabetic Neuropathy** | | | | | |
| **Benson [89]** | DL, in conjunction with TL | VGG16 CNN | retinal changes as a predictor of diabetic neuropathy | Accuracy | 89% |
| | | | | Sensitivity | 78% |
| | | | | specificity | 95% |
| **Cervera [90]** | DL | SqueezeNet, Inception, and DenseNet | retinal changes as a predictor of diabetic neuropathy | AUC (95% CI) | 0.8013 (±0.0257) |
| **Diabetic Nephropathy** | | | | | |
| **Sabanayagam [101]** | DL | CondenseNet | Image only showing retinal changes as a predictor of diabetic nephropathy | AUC (95% CI) | 0·911 (0·886–0·936) |
| | | | | Sensitivity | 0·83 |
| | | | | Specificity | 0·83 |
| | | | RF only includes age, sex, ethnicity, diabetes, and hypertension | AUC (95% CI) | 0·916 (0·891–0·941) |
| | | | | Sensitivity | 0·82 |
| | | | | Specificity | 0·84 |
| | | | hybrid DLA combining image and RF | AUC (95% CI) | 0·938 (0·917–0·959) |
| | | | | Sensitivity | 0·84 |
| | | | | Specificity | 0·85 |
| **Zhang [100]** | DL (CNN) | ResNet-50 | Image only showing retinal changes as a predictor of diabetic nephropathy | AUC | 0.829-0.918 |
| | | | RF only (age, sex, height, weight, body-mass index, and blood pressure) | AUC | 0.787-0.861 |
| | | | hybrid DLA combining image and RF | AUC | 0.845-0.930 |
| **Diabetic peripheral arterial disease** | | | | | |
| **Mueller [107]** | DL (CNN) | Multiple Instance Learning (MIL) | retinal changes as a predictor of diabetic PAD | Accuracy | 0.674- 0.837 |
| | | | | ROC AUC | 0.653-0.890 |

**3.3 AI techniques used in retinal images analysis**

AI researchers have used various models to analyze retinal images; the two most common neural are the CNN-based architectures ResNet [109-111] and VGGNet [112, 113]. Figure 5 illustrates the architecture of a CNN specifically designed for image analysis. Deep networks often falter in practice due to difficulty optimizing the networks caused by the vanishing gradient problem. ResNet circumvents this issue by residual learning, which uses skip or shortcut connections to skip one or more layers during forward and backward passes [109, 114]. This ensures that the deeper layers produce no higher training errors than their shallower counterparts, which makes it easier for the network to learn identity functions. ResNet's modular architecture also facilitates up or down scaling of the model by adjusting the number of residual blocks. This makes it an ideal choice for developers, as they can tailor the architecture to specific needs and computational constraints. Belying its depth, ResNet consumes comparatively fewer computational resources, which underscores its efficiency. VGGNet is a family of deep CNNs distinguished by its simple design, ability to capture complex image characteristics, and excellent image classification performance. VGG16, a notable variant comprising 16 layers [112, 113], begins with layers containing 64 channels, with a 3x3 filter size and consistent padding, which are succeeded by a max-pooling layer with a stride of (2, 2), and subsequent convolution layers that progressively increase in channels, to for example 128, while maintaining a uniform 3x3 filter size. VGGNet is predominantly engineered for image classification on extensive datasets, notably the ImageNet database, which encompasses over 14 million images categorized based on the WordNet structure. Beyond image classification, advancements in VGG architectures have been pursued to cater to diverse computer vision applications and to augment its efficacy in classification problems [115-117].

Inception-v3, DenseNet, AlexNet, ResNet, and U-Net are neural network architectures with distinct structural variations that have been applied to retinal image analysis [118, 119]. Inception-v3, from the GoogLeNet family, was designed for multi-level feature extraction tailored to large-scale image recognition tasks, incorporating factorization and other approaches, such as batch normalization, to minimize parameters and maximize performance [120, 121]. DenseNet is well-suited for image classification and segmentation due to dense connections between layers, which improve gradient flow and encourage feature reuse [122]. AlexNet, an early CNN, is adept at image classification due to its deep architecture and use of the rectified linear unit (RELU) activation but may be outperformed by later models in terms of efficiency [123]. By expanding on residuals and using split-attention methods, ResNet is able to train the model to focus on the most relevant characteristics for image classification [124]. Unlike Inception-v3 and AlexNet, which are more general-purpose models, U-Net was developed specifically for segmenting biological images [125]. Its "U-shaped" design allows for more accurate localization, which is important in diagnostic imaging [126].

Supervised ML learns by mapping input signals or images to their respective labels. In retinal image analysis, supervised ML techniques have been extensively employed for the detection and classification of various retinal diseases, including DR. With labeled retinal images, the algorithms can effectively learn the patterns associated with DR, thereby facilitating early diagnosis and timely interventions to avert irreversible vision loss [23, 127]. Transfer learning leverages learning gained from vast datasets, such as ImageNet [128], circumventing the need for vast retina-specific training datasets [129]. For retinal image analysis, the pre-trained model's final layers are typically adjusted to cater to the specificities of retinal diseases or features. After this adjustment, the model is fine-tuned on available retinal datasets. Transfer learning accelerates the training process, often yielding models with enhanced accuracy and robustness that can be applied to diverse tasks, including DR detection, retinal lesion identification, and retinal vessel segmentation [23, 50].

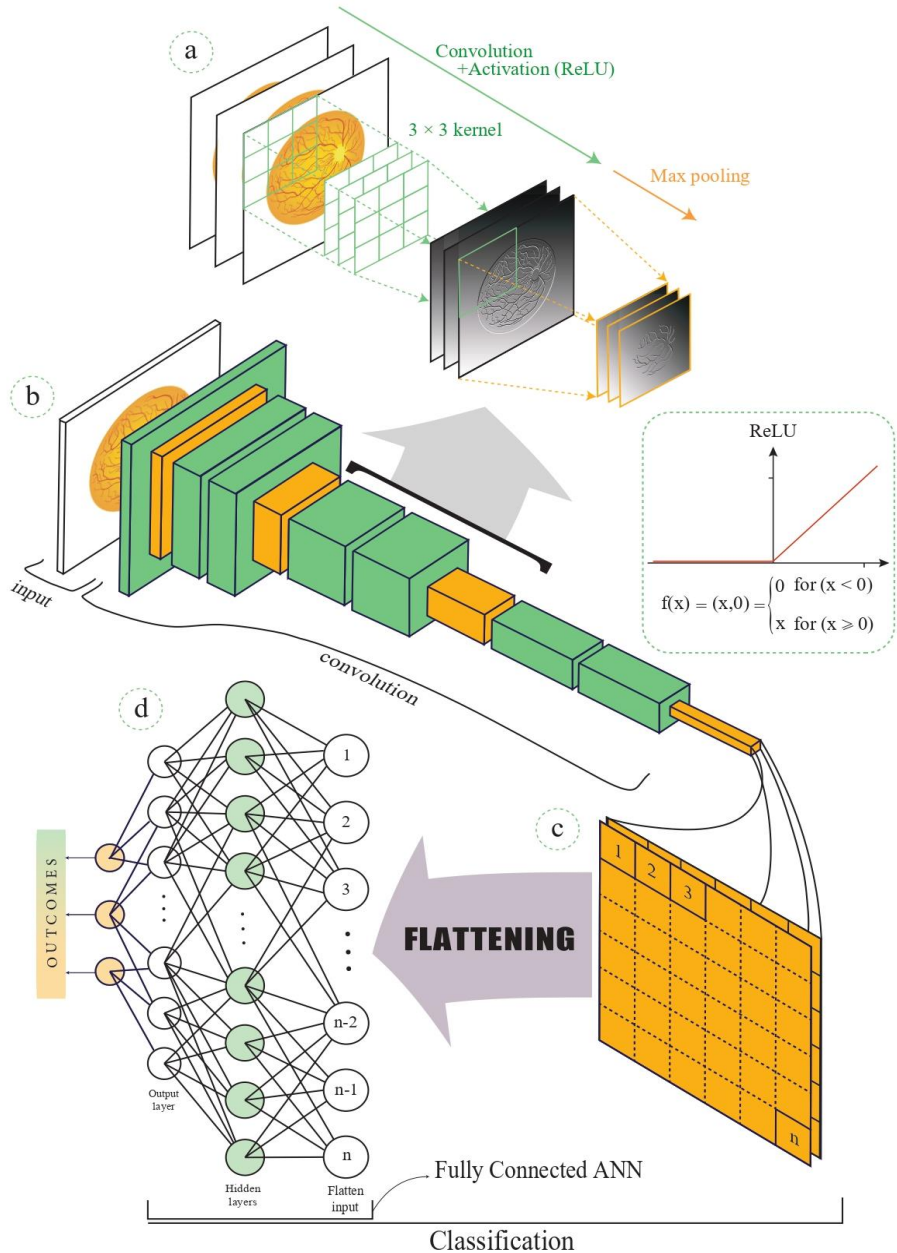

**Figure 5**. A detailed look at the architecture of a convolutional neural network (CNN) for image analysis. a: The convolutional layers (green) use kernels that slide over the three-channel RGB image to recognize key features from the input image. Following the convolution process, the Rectified Linear Unit (ReLU) activation function (green) is applied to introduce non-linearity, enhancing the network's capability to learn intricate patterns. Subsequently, the max pooling process (orange) is applied, reducing the spatial dimensions by selecting the maximum value within specified regions. **b:** this part is the multiple repetitions of convolution, ReLU activation function, and max pooling processes, creating the final feature map. **c:** shows flattening of the last max pooling layer, which converts the 2D feature maps into a 1D vector to prepare the data for the upcoming fully connected layers. **d:** the architecture transitions to fully connected layers, leading to a classification process where the features are used to provide definitive conclusions.

## 4. Challenges of AI in diabetes care

Only two Asian countries have DR screening programs that conform to International Council of Ophthalmology standards [130]. There is not only an imperative to standardize national protocols to mitigate gaps in screening and referral timelines [131] but also a need to establish comprehensive guidelines for AI implementation in DR screening to ensure standardized and effective practice [132]. AI-enabled DR screening can reduce economic constraints and enhance accessibility to healthcare services [133], but several obstacles remain to surmount [134]. Overcoming these challenges will require multidisciplinary cooperation, data standardization, resource sharing, real-world verification, and productization [135]. In particular, DL algorithms require large datasets with thousands or millions of images for training, which are costly to label and curate. AI developers often resort to using available but limited training datasets, which may not be generalizable to the real world, where image quality may be affected by deficiencies in the clinical setting and are not of consistently high quality [136]. Suboptimal image quality and low target pathology incidence may drive higher false-positive rates [137]. Even if image issues are resolved, AI algorithms that only focus on retinal image assessment will not be able to analyze the clinical and psychosocial aspects that can modulate the diagnosis, perception, coping mechanisms, and holistic management of individual patients [138]. The lack of compatibility among electronic health record vendors impedes such clinical data integration and may delay the execution of computer-interpretable guidelines for clinical decision-making in DR [139]. To garner wider adoption, there is a need for more research on the assessment and testing of AI diagnostic tools in clinical settings [140], even as validating AI algorithms in different populations and camera systems remains challenging [141]. Additionally, conducting rigorous clinical trials of AI models, particularly randomized controlled trials, requires meticulous methodological adjustments and considerations of clinical equipoise, informed consent, and fairness [142]. Practical challenges to building and deploying AI at scale include regulatory pressures, conflicting business goals, and data quality issues [143]. Even as AI promises solutions to problems through technological advances, AI in healthcare raises ethical and legal issues [144]. As AI systems , become more autonomous [145] the need to incorporate ethical considerations and moral reasoning into model decision-making processes has become more pressing. Equally important is the obligation to identify and resolve ethical issues—inclusivity, bias, and social acceptability—to ensure fair user access to promising healthcare AI technologies [146]. Issues such as protecting sensitive medical information during image analysis or remote sharing via tele-retinal screening [147] also raise patient privacy concerns, requiring careful system security during AI solution development and deployment.

Current approaches to AI model design are delinked mainly from the complex healthcare environments they are intended for; as a result, the development of AI models has vastly outpaced their adoption into existing clinical workflows [148]. This separation contributes to models that lack clear use cases and are neither tested nor scaled in clinical settings. A mixed-methods approach that integrates design thinking and quality improvement methodologies—aiming to understand variations in healthcare processes and incorporating user-centered design to ensure model functionality in practice [149]—can potentially redress this gap to smoothen AI integration within the healthcare domain [150], and garner wider clinical adoption [151].

AI screening systems have shown promising results in detecting DR from CFP and OCT images. However, there is still a need for validation, regulation, safe implementation, and demonstration of clinical impact before widespread adoption. Test-bedding new AI models in clinical settings is essential for identifying and remedying system bugs before full-scale deployment [152, 153]. Further, the AI tools may need to be validated and calibrated against local populations and clinical contexts: published results from one context may not always be generalizable or achievable in a different setting [154]. Finally, in complex AI models of DR diagnosis or treatment decision-making, interpretability is essential for gaining the trust of clinicians and patients alike. There is a need for transparent and interpretable AI in the reasoning processes behind the generated model outputs. Explainable AI is an active area of research and remains a challenge [155].

5. Potential enhancements and future directions of AI in DR management

At the intersection of telemedicine and AI, tele-retinal image analysis promises to democratize access to screening and downstream healthcare services, transform the management of DR, and improve long-term patient outcomes while reducing financial and time costs for both patients and payers. AI-enabled DR screening has demonstrated encouraging outcomes, with DL algorithms yielding high levels of sensitivity and specificity. The operational efficiency of community-based tele-retinal image analysis may be enhanced: the gradability of retinal images can be assessed at source, expediting identification of poor-quality images for either manual or additional AI-based grading (the ultimate choice will depend on the associated labor cost vs.

intrinsic diagnostic value-add of AI) [156-158]. Future technological developments in AI can introduce significant opportunities for technical refinements to optimize DR diagnosis and downstream management. Faster AI software processing speeds will enhance system accuracy and efficiency, facilitate seamless and responsive navigation of the AI interface by clinician users, and enable more effective preventive therapeutic interventions. Independent of these developments, real-world implementation of AI technology presents its practical challenges, e.g., workflow integration, technical adaptability, ethical implications, cost-effectiveness considerations, etc. To ensure the ethical and balanced integration of AI into DR screening programs, the governance model for AI implementation must focus on honesty, equality, reliability, and responsibility [152-154].

Explainable AI—AI systems that provide understandable explanations for their decisions and actions—make the complex decision-making processes of AI models transparent and interpretable, which is essential, given the clinical impact of healthcare decisions, for gaining the trust and acceptance of AI model outputs by doctors and patients. The applications of explainable AI in the medical domain are vast and transformative [159, 160], encompassing diverse tasks, such as decision-making, risk management, predictions, and medical image analysis (the sensitivity of AI for detecting abnormalities often surpasses that of the human eye). Explainable AI aids in explaining AI-driven insights. For example, while a traditional AI model may provide a diagnosis with a confidence score ("Diabetic retinopathy detected with 95% confidence"), ophthalmologists may hesitate to Google Net trust the result or administer recommended treatment without knowing the basis for the diagnosis. With explainable AI, the system can highlight regions in the retina or specific lesions, like hemorrhages or microaneurysms, critical to the model decision via heat maps. These allow clinicians to relate to the results like traditional funduscopic examination, and the maps may additionally serve as useful guides to the planning of photocoagulation therapy, where applicable [161, 162].

Cloud-based systems revolutionize diabetes management and prevention by enhancing data processing, enabling real-time interventions, and optimizing resources. They facilitate early detection, personalized therapy, and swift glucose fluctuation response. Migration to cloud structures reduces costs and administrative burdens, while user-friendly digital platforms support self-monitoring and community engagement. Integrating artificial intelligence with cloud platforms promises sharper insights for combating diseases like Type II diabetes, leading to societal benefits. Moreover, some researchers have investigated the potential of cloud-based systems for diabetes management and prevention [163, 164]. A study by Salari et al. [165] has shown promising results in using mobile and cloud systems to improve self-care for chronic conditions, offering hope for more effective and user-friendly solutions. Similarly, R. Nasser et al. [166] have proposed innovative methods using advanced technology, such as AI and cloud computing, to predict glucose levels and integrate them with wearable devices. These advancements point toward a future where cloud-based systems could revolutionize how diabetes is managed, offering personalized and timely interventions to enhance health outcomes for individuals with the condition. Figure 6 illustrates the interconnected components of a cloud-based system for diabetic retinopathy screening.

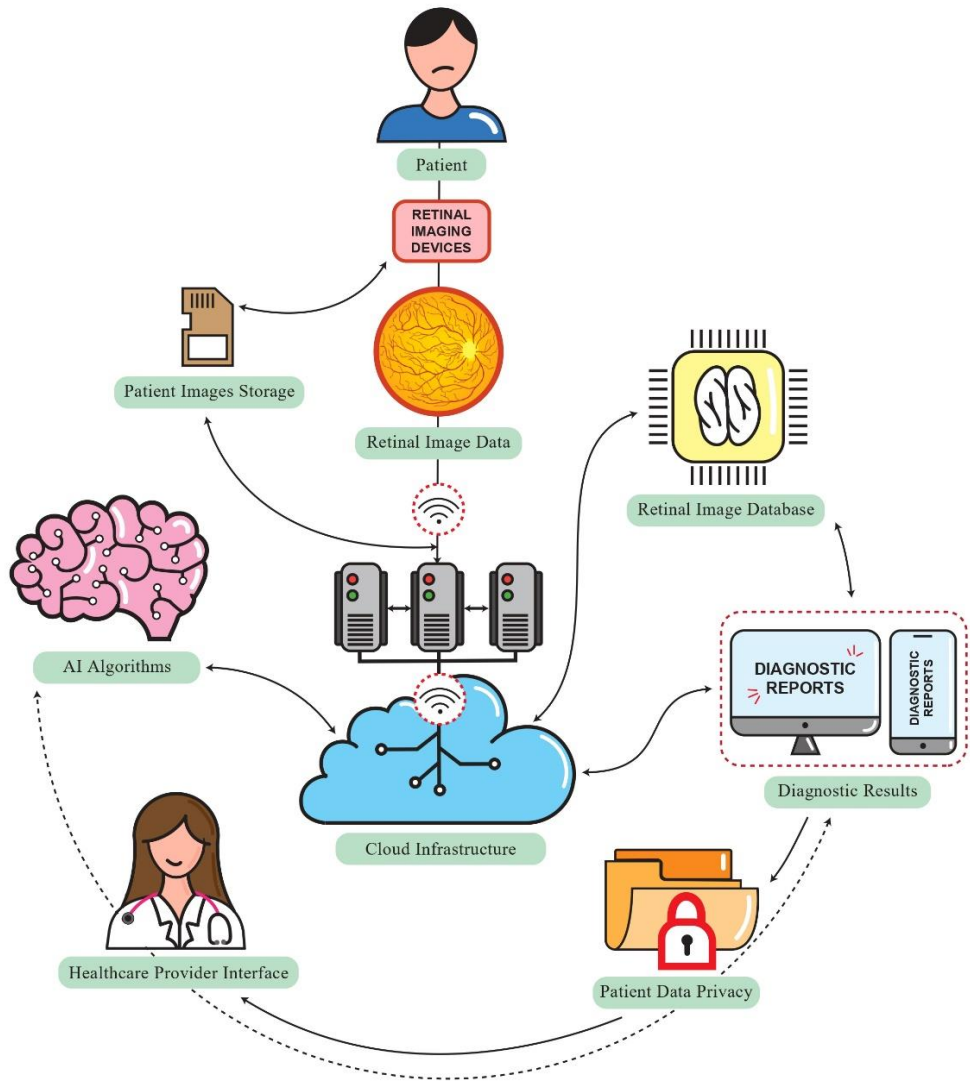

**Figure 6.** This schematic illustrates the interconnected components of a cloud-based system for diabetic retinopathy screening. The cloud infrastructure, depicted as remote servers and databases, processes retinal image data captured by a retinal imaging device, such as a fundus camera or OCT scanner. AI algorithms analyze the images within the cloud, generating diagnostic results like risk scores for diabetic retinopathy. These results are accessible to healthcare providers through interfaces on computers or mobile devices, ensuring prompt patient care. Patient data privacy measures safeguard sensitive information, including encryption and secure transmission protocols. Additionally, a feedback loop may exist, where diagnostic results contribute to the continuous improvement of AI algorithms over time.

## 6. Conclusion

Given the increasing prevalence and incidence of DM, developing cost-effective, population-based DR screening and management strategies is becoming increasingly important nationally. AI has shown significant promise in accurately diagnosing DR, with potential applicability extending beyond the diagnosis and grading of DR. This includes the diagnosis of diabetic neuropathy, diabetic nephropathy, and cardiovascular diseases. These applications may involve using CFP independently or in conjunction with other advanced diagnostic techniques, such as OCT and various clinical parameters. AI's capacity to assess a patient's health about DM complications and to forecast the risk of future complications positions AI-assisted retinal image analysis as a potentially key component in personalized medicine for individuals with DM.

**Declaration**

**Ethics approval and consent to participate:**

This study was conducted under the principles outlined in the Declaration of Helsinki.

**Consent for publication**

N/A

**Availability of data**

N/A

**Competing interests**

The authors declare that there is no conflict of interest.

**Funding Sources**

None.

**Authors' contributions**

AJ conceptualized the manuscript. YSB, MM, MA, AJ, and NS conducted a literature review. YSB, MM, MA, and NS drafted the manuscript. SP, RST, RA, URA, SMSI, NS, MA, and AJ provided comments and revised the manuscript. NS and AJ created and designed all the figures. All authors read and approved the final version of the manuscript. It should be noted that YSB, MM, and NS contributed equally to this study, and AJ and RA are co-corresponding authors of this manuscript.

**Acknowledgments**

None.